\documentclass[aps,reprint,pre]{revtex4-1}
\usepackage{graphicx}
\usepackage{subfigure}
\usepackage[table,dvipsnames]{xcolor}
\usepackage{longtable}
\usepackage{multirow}
\usepackage{epstopdf, epsfig}
\usepackage{amsmath}
\usepackage{amsfonts}
\usepackage{dcolumn}
\usepackage{bm}
\usepackage{color, colortbl}
\definecolor{LightGray}{rgb}{0.88,0.88,0.88}

\setcitestyle{square}

\renewcommand{\Re}{\operatorname{Re}}
\renewcommand{\Im}{\operatorname{Im}}

\begin{document}

\title{Enhanced Tuneable Rotatory Power in a Rotating Plasma}
\author{Renaud Gueroult}
\affiliation{LAPLACE, Universit\'{e} de Toulouse, CNRS, INPT, UPS, 31062 Toulouse, France}
\author{Jean-Marcel Rax}
\affiliation{Universit\'{e} de Paris XI - Ecole Polytechnique, LOA-ENSTA-CNRS, 91128 Palaiseau, France}
\author{Nathaniel J. Fisch}
\affiliation{Department of Astrophysical Sciences, Princeton University, Princeton, NJ 08540, USA}
\date{\today}

\begin{abstract}
The gyrotropic properties of a rotating magnetized plasma are derived analytically. Mechanical rotation leads to a new cutoff for wave propagation along the magnetic field and polarization rotation above this cutoff is the sum of the classical magneto-optical Faraday effect and the mechanico-optical polarization drag. Exploiting the very large effective group index near the cutoff, we expose here, for the first time, that polarization drag can be $10^4$ larger than Faraday rotation at GHz frequency. The rotation leads to weak absorption while allowing direct frequency control, demonstrating the unique potential of rotating plasmas for non-reciprocal elements. The very large rotation frequency of a dense non-neutral plasma could enable unprecedented gyrotropy in the THz regime. 
\end{abstract}

\maketitle

\paragraph*{Introduction.} Non-reciprocity, that is the property in certain electromagnetic systems that the field produced by a given source changes if the source and the detector are interchanged, underpins numerous fundamental phenomena (see, \emph{e.~g.}, Refs. ~\cite{Potton2004,Drezet2014,Caloz2018}). Non-reciprocal (NR) signal processing elements such as circulators, gyrators and isolators are also essential for applications where one-way propagation is required. They are notably used in optics to eliminate cross-talk and feedback in lasers~\cite{Petermann1995} and telecommunication networks~\cite{Shirasaki1982,Bi2011,Fan2012}, at RF frequencies to enable simultaneous emission and reception from a single antenna (in-band full-duplex wireless)~\cite{Reiskarimian2016}, and at microwave frequencies to handle load reflection in thermonuclear fusion high power heating systems~\cite{Dixit2017}.

The development of NR elements has largely relied on the magneto-optical (also known as gyrotropic) properties of ferrites (a non-conductive ferrimagnetic material). In a static magnetic field, ferrites' permeability tensor is non-diagonal at microwave frequencies due to the precession of electrons spin~\cite{Hogan1953,Adam2002} while ferrites' permittivity tensor is non-diagonal at optical wavelengths as a result of spin-orbit coupling~\cite{Dionne2005}. Non-zero off-diagonal terms then lead to the well known Faraday rotation~\cite{Faraday1846}, that is to a non-reciprocal rotation of the polarization of waves propagating along the static magnetic field. These two distinct phenomena and the associated Faraday rotation are respectively the basis for microwave~\cite{Hogan1952,Morris1957} and optical~\cite{Aplet1964} ferrite isolators. However, ferrite-based NR elements suffer from shortcomings. First, the NR properties of ferrite-based elements are hardly tunable. For instance, the length of an isolator is chosen such that a $45^{\circ}$ polarization rotation is obtained at a given operating magnetic field and for a given wavelength. Second, the intrinsic losses of ferrites at millimeter and sub-millimeter wavelengths makes them inapplicable in the rapidly growing terahertz regime. Despite new ferrite treatments~\cite{Yu2018} or specifically developing ceramics~\cite{Yu2017}, the issue of tunability remains. These intrinsic limitations of ferrites have motivated the search for alternative NR materials. 

An often sought out property in this search for NR materials is enhanced gyrotropy. In natural media, both the electron cyclotron frequency and electron spin precession frequency are typically in the microwave range so that magneto-optical effects tend to be weak at optical frequencies. Enhanced effects make it possible to shorten the propagation length required to yield a given polarization rotation, and are thus key for the miniaturization of these devices such as needed for integrated optics~\cite{Dai2012}. Another rationale for enhanced effects is that the shorter the propagation length, the smaller the losses. Enhanced gyrotropy therefore holds the promise for more compact and lower-losses devices.

Among the candidates are meta-materials (MMs) designed to exhibit gyrotropic properties~\cite{Kodera2011,Kalev2013}, which can yield extremely strong rotatory power (polarization rotation per unit length)~\cite{Kuwata-Gonokami2005,Rogacheva2006} and offer great opportunities for light polarization manipulation~\cite{Cong2012,Floess2015,Firby2016}. However, while some effort has been made to develop MMs with tunable gyrotropic properties~\cite{Floess2015,Zhu2015}, MMs' properties are basically set by design. Another possibility is magneto-statically biased graphene~\cite{Crassee2010,Sounas2011,Tamagnone2016}, where the direction of polarization rotation can be controlled through a change in the sign of charged carriers induced by an electric bias~\cite{Sounas2012,Poumirol2017}. However, because graphene is conductive at microwave frequencies, the promise of tunable graphene-based NR elements is offset by comparatively larger losses.

Another route to create NR elements is moving media~\cite{Jackson1970}. In the laboratory frame, the permittivity tensor of a rotating isotropic dielectric is non-diagonal~\cite{Gueroult2019a}, leading to a NR polarization rotation of a wave propagating along the rotation axis~\cite{Fermi1923,Player1976}. This phenomena is sometimes referred to as polarization drag or mechanical Faraday rotation. In typical dielectrics, the associated specific rotatory power $\delta\sim10^{-5}$~rad~m$^{-1}$ is extremely small~\cite{Jones1976}, which strongly limits its potential for applications. However, enhancements by a factor $10^5$ can be achieved by exploiting the very large effective group index $n_g$ produced by resonant conditions in a ruby window~\cite{Franke-Arnold2011}, or the extreme angular rotation frequency of a gas of molecular super-rotors~\cite{Steinitz2020}.

Here, we identify for the first time an enhanced gyrotropy effect by mechanical rotation in a plasma. For a modest rotation frequency of a $100$~Hz the specific rotatory power of a rotating magnetized plasma~\cite{Lehnert1971} for GHz frequency waves can be $10^4$ larger than that obtained from the magneto-optic polarization rotation alone absent rotation. Both the frequency and its strength can be tuned by modifying plasma parameters. The comparatively weak losses found in rotating plasmas suggest unprecedentedly high mechanico-optical figure of merit. Importantly, the fast rotation of a dense non-neutral plasma enables enhanced gyrotropy in the THz regime.

\paragraph*{Mechanical polarization rotation in a plasma.} Let us consider a cold and collisionless magnetized plasma in rigid body rotation such that the plasma angular frequency $\bm{\Omega}$ and the static magnetic field $\mathbf{B}_0$ are parallel to the waves' propagation direction $\mathbf{\hat{z}}$. Assuming that the dielectric properties in the medium's local rest frame are not modified by rotation, the wave index for right- and left-circularly polarized (RCP and LCP) waves propagating along $\mathbf{\hat{z}}$ write (see Ref.~\cite{Gueroult2019a})
\begin{align}
{n_{rcp/lcp}}^2(\omega) & = 1+ \bar{\chi}_{\perp}(\omega') \pm \bar{\chi}_{\times}(\omega')\nonumber\\
 & \qquad -\frac{\Omega}{\omega}\left[\bar{\chi}_{\times}(\omega') \pm \bar{\chi}_{\parallel}(\omega')\pm\bar{\chi}_{\perp}(\omega')\right],
\label{Eq:index_general}
\end{align} 
with $\bar{\chi}_{\parallel}(\omega) = -\sum{\omega_{p\alpha}}^2/\omega^2$, $\bar{\chi}_{\perp}(\omega)=\sum{\omega_{p\alpha}}^2/[{\omega_{c\alpha}}^2-\omega^2]$ and $\bar{\chi}_{\times}(\omega) = \sum\varepsilon_{\alpha}\omega_{c\alpha}{\omega_{p\alpha}}^2/[\omega(\omega^2-{\omega_{c\alpha}}^2)]$ the standard components of the plasma susceptibility tensor $\bar{\bm{\chi}}$ in the plasma rest frame and $\omega' = \omega\mp\Omega$ the Doppler shifted wave angular frequency. Here the sums run on all plasma species $\alpha$, $\omega_{p\alpha}=[n_{\alpha}e^2/(m_{\alpha}\epsilon_0)]^{1/2}$ and $\omega_{c\alpha} = q_{\alpha B_0/m_{\alpha}}$ are the plasma and cyclotron angular frequencies, respectively, and $\varepsilon_{\alpha} = q_{\alpha}/|q_{\alpha}|$. For a collisionless plasma, the susceptibility is real and propagation is lossless.

The difference in wave index for LCP and RCP waves $\Delta n (\omega) = n_{lcp}(\omega)-n_{rcp}(\omega)$ leads to circular birefringence with a change in polarization angle $\phi$ (specific rotatory power) governed by
\begin{equation} 
\delta = \frac{d \phi(\omega)}{dz}  = \frac{\Delta n(\omega)}{2}\frac{\omega}{c}.
\label{Eq:general_polarization_rotation}
\end{equation}
Absent rotation ($\Omega=0$), Eq.~(\ref{Eq:index_general}) yields the classical magneto-optical Faraday effect, and polarization rotation stems entirely from to the non-diagonal components $\bar{\chi}_{\times}$ of the rest-frame susceptibility tensor. However, since $|\bar{\chi}_{\parallel}|\propto\omega^{-2}$ Eq.(\ref{Eq:index_general}) shows that there is a cutoff frequency $\omega_c$ below which one of the CP modes does not propagate when $\Omega\neq0$~\cite{Gueroult2019a}. Above this mechanically induced cutoff Faraday rotation is supplemented by polarization drag, whereas none of these effects is found below $\omega_c$. Assuming $B_0>0$, the cutoff is for the LCP (RCP) wave if $\Omega>0$ ($\Omega<0$), in which case polarization drag subtracts from (adds to) Faraday rotation. For simplicity, we consider $\Omega>0$ and $B_0>0$.  

\begin{figure}[htbp]
\begin{center}
\includegraphics[]{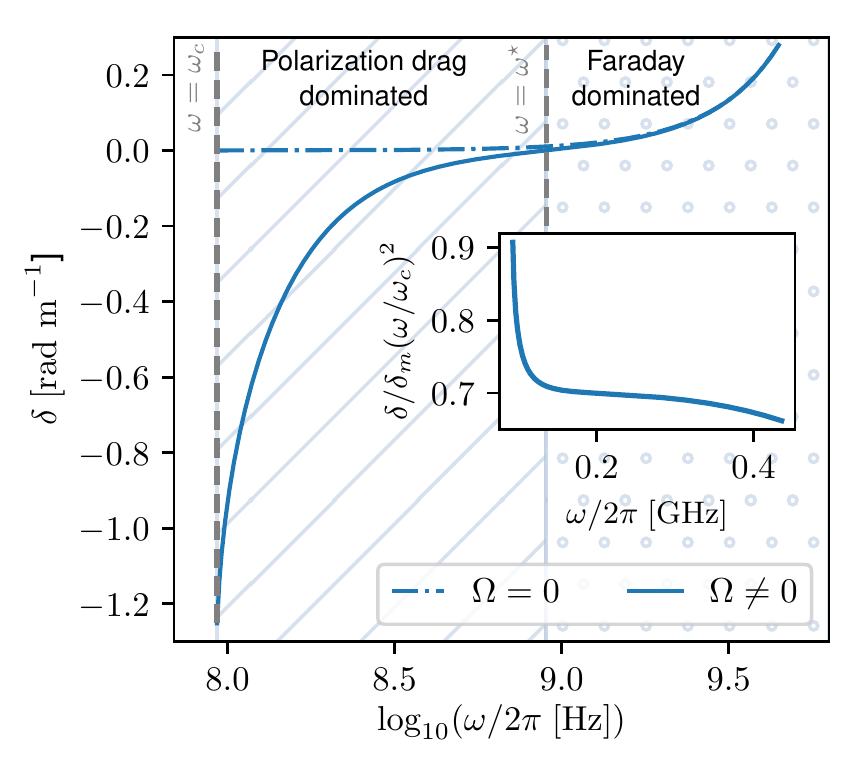}
\caption{Polarization rotation by unit length of propagation with and without rotation. Conditions correspond to the baseline parameters given in Table~\ref{Tab:params}. A sign reversal and amplification is observed below the crossover frequency $\omega^{\star}$ in the presence of mechanical rotation ($\Omega\neq0$). The ion cyclotron frequency $\omega_{ci}/(2\pi)\sim15$~GHz is above the frequency band considered here. The inset highlights the $\delta\propto-\omega^{-2}$ high frequency scaling of mechanical polarization rotation. }
\label{Fig:Fig1}
\end{center}
\end{figure}

Eqs.~(\ref{Eq:index_general}) and (\ref{Eq:general_polarization_rotation}) show that polarization drag can actually dominate over the intrinsic Faraday rotation if the part of the mechanical contribution to wave indexes that differs for RCP and LCP waves, that is $|\Omega\left[\bar{\chi}_{\parallel}(\omega')+\bar{\chi}_{\perp}(\omega')\right]/\omega|$, is greater than the non diagonal term $|\bar{\chi}_{\times}(\omega')|$. From the definition of $\bar{\bm{\chi}}$, this condition is always met for low enough wave frequency since $|\bar{\chi}_{\parallel}|\propto\omega^{-2}$ while $|\bar{\chi}_{\times}(\omega)|\propto\omega$ for $\omega\ll\omega_{ci}$ (see Ref.~\cite{GueroultSupp} for the details on low frequency dielectric properties of a rotating plasma). Assuming a slowly rotating ($\Omega\ll\omega$) and underdense plasma ($\omega_{ce}\gg\omega_{pe}$) so that $|\bar{\chi}_{\parallel}(\omega)|\gg\bar{\chi}_{\perp}(\omega)$, then 
\begin{equation}
\delta \sim \frac{\omega}{2nc}\left[-\bar{\chi}_{\times}(\omega)+\frac{\Omega}{\omega}\bar{\chi}_{\parallel}(\omega)\right]
\end{equation}
to lowest order in $\Omega/\omega$ above the cutoff with $n = (n_{lcp}+n_{rcp})/2$. The first and second term in the right hand side bracket are respectively the classical Faraday effect and the polarization drag. The crossover frequency for which these two effects have comparable and opposite amplitudes then simply writes~\cite{GueroultSupp}
\begin{equation}
\omega^{\star} \sim\eta[\Omega{\omega_{ce}}^3]^{1/4}\propto {\Omega}^{1/4}{B_0}^{3/4}
\label{Eq:crossover}
\end{equation} 
with $\eta^2$ the electron to ion mass ratio. Below this frequency and down to the cutoff frequency 
\begin{equation}
\omega_c \sim [{\omega_{pe}}^2\Omega]^{1/3} \propto {\Omega}^{1/3}{n_e}^{1/3},
\label{Eq:cutoff}
\end{equation}
mechanical polarization rotation dominates over Faraday rotation. This behaviour is confirmed when solving numerically $\delta$ from Eqs.~(\ref{Eq:index_general}), as illustrated in Fig.~\ref{Fig:Fig1} for the baseline parameters set (${n_e}^{\diamond}$, ${B_0}^{\diamond}$, ${\Omega}^{\diamond}$) given in Table~\ref{Tab:params}. Importantly, note that the particular choice here of ${B_0}^{\diamond} = 10^3$~T (consistent with state of the art laser-driven capacitor-coil target experiments~\cite{Santos2015}) is only dictated by the choice of a cutoff of $0.1$~GHz, but that this effect can in principle be observed for much weaker magnetic fields, albeit at lower wave frequency.

Eqs.~(\ref{Eq:crossover}) and (\ref{Eq:cutoff}) reveal two of the promising tunability features of polarization drag in a plasma. First, the frequency band $[\omega_c,\omega^{\star}]$ over which polarization drag dominates can be broadened by increasing the magnetic field $B_0$ and, to a lesser extent, the rotation frequency $\Omega$. Second, this frequency can be upshifted by increasing the plasma density $n_0$ or the mechanical rotation frequency $\Omega$.  

\begin{table}[htbp]
\begin{center}
\caption{Baseline plasma parameters and corresponding circular birefringence properties}
\label{Tab:params}
\begin{tabular}{c c c }
\hline
\hline
Plasma density & ${n_e}^{\diamond}$ & $10^{20}$~m$^{-3}$\\
Magnetic field & ${B_0}^{\diamond}$ & $10^3$~T\\
Rotation frequency & $\Omega^{\diamond}$ & $100$~Hz\vspace{0.1cm}\\
Cutoff frequency & $\omega_c$ & $0.1$~GHz\\
Crossover frequency & $\omega^{\star}$ & $1$~GHz\\
Enhancement factor & $\gamma$ & $10^4$\\
\hline
\hline
\end{tabular}
\end{center}
\end{table}

\paragraph*{Enhanced gyrotropy.} Taylor expanding Eq.~(\ref{Eq:index_general}) shows that the wave index difference $\Delta n = n_l-n_r=-\sqrt{2}+\sqrt{3\varpi}$ to lowest order in $\varpi = \omega/\omega_c-1$. This implies that the specific rotatory power $\delta$ is maximum at the cutoff with 
\begin{equation}
\delta_{m} \doteq |\delta(\omega_c)|\sim\frac{\omega_c}{\sqrt{2}c}.
\label{Eq:rotatory_power}
\end{equation}
Above the cutoff but below the crossover frequency $\omega^{\star}$, $|\delta|$ first decreases as $1-\sqrt{3\varpi/2}$ while $\varpi\ll1$ and then follows the high-frequency asymptotic behaviour for mechanical polarization rotation $\delta\propto -\Omega\omega_{pe}/\omega^2$ as can be seen in the inset in Fig.~\ref{Fig:Fig1}. 

To highlight the enhancement produced by mechanical rotation,  $\delta_{m}$ is to be compared with the specific rotatory power absent of rotation $\delta_{\Omega=0}(\omega_c)$. Since $\bar{\chi}_{\times}(\omega) \sim -{\omega_{pi}}^2\omega/{\omega_{ci}}^3$ for $\omega\ll\omega_{ci}$, one gets 
\begin{equation}
\gamma\doteq\frac{\delta_{m}}{\delta_{\Omega=0}(\omega_c)}\sim \frac{1}{\sqrt{2}}\left(\frac{{\omega_{ci}}}{\omega_{pi}}\right)^2\frac{{\omega_{ci}}}{\omega_c}\propto \frac{\eta^{2}{B_0}^{2}}{{\Omega}^{1/3}{n_e}^{4/3}}.
\label{Eq:gamma}
\end{equation}
For the baseline parameters (${n_e}^{\diamond}$, ${B_0}^{\diamond}$, ${\Omega}^{\diamond}$) (see Table~\ref{Tab:params}), this yields $\gamma\geq10^4$ which is comparable to the enhancement obtained by exploiting resonant conditions in a ruby window~\cite{Franke-Arnold2011}, or that of a gas of molecular super-rotors~\cite{Steinitz2020} but for a mechanical rotation frequency $10$ orders of magnitude smaller. Note that this circular birefringence enhancement in a rotating plasma near the cutoff frequency can be interpreted, similarly to the enhancement found using slow light in a ruby window~\cite{Franke-Arnold2011}, as the effect of a very large effective group index. Indeed, the group velocity $d\omega/dk$ of the LCP wave tends to zero as $\omega$ approaches $\omega_c$~\cite{GueroultSupp}. 

Since $\delta_m\propto\omega_c$, the specific rotatory power can also be controlled online through the rotation frequency $\Omega$ and the plasma density $n_e$. Quantitatively, $\delta_{m}\sim1$~rad~m$^{-1}$ for the baseline parameters (${n_e}^{\diamond}$, ${B_0}^{\diamond}$, ${\Omega}^{\diamond}$), which is again comparable to that obtained by exploiting slow light conditions in a ruby window~\cite{Franke-Arnold2011}, or the extremely large rotation of molecular super-rotors~\cite{Steinitz2020}. Also, while such specific rotatory power values are well below those obtained at THz frequency for instance on epitaxial HgTe thin films ($\sim10^6$~rad~m$^{-1}$)~\cite{Shuvaev2011} or in graphene ($\sim10^8$~rad~m$^{-1}$)~\cite{Crassee2010}, magnetized plasma channels' length can be much longer than the thickness $d$ of these media. In terms of achievable rotation angle $\theta = \delta d$, the weaker specific rotatory power in plasmas is thus partly offset by a longer propagation length. Quantitatively, one gets $\theta\sim10^{-3}$~rad for a mm long plasma channel~\cite{Santos2015} with parameters (${n_e}^{\diamond}$, ${B_0}^{\diamond}$, ${\Omega}^{\diamond}$) versus $\theta\sim1$~rad in epitaxial HgTe thin films or graphene, and the frequency scaling displayed in Eq.~(\ref{Eq:rotatory_power}) suggests that this gap would be further reduced for higher frequency operation.


\paragraph*{Low loss.} In NR components, losses are often characterised by the absorption coefficient $\alpha(\omega)$ which is related to the transmission $T$ through $T = \exp[-\alpha(\omega)d]$, with $d$ the propagation length. In many NR media, losses can be significant. In permanent magnets for instance $\alpha$ grows with $\omega$ and $\alpha d\geq1$ for $\omega/(2\pi)\geq 0.3$~THz~\cite{Shalaby2013}.  Similarly, the structure of graphene metasurface can be designed to maximise transmission at a given frequency, but transmission drops drastically out of this frequency band ($T<0.3$ for $|\Delta \omega|/\omega\geq1\%$)~\cite{Qin2018}.

\begin{figure}
\begin{center}
\includegraphics[]{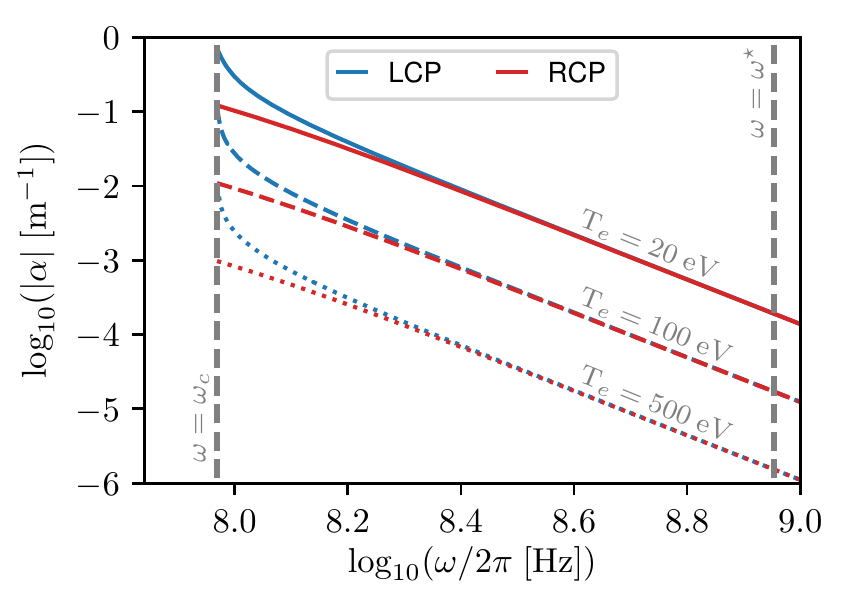}
\caption{Absorption coefficient $\alpha = 2 \omega \Im[n(\omega)]/c$ for RCP and LCP normal modes as a function of frequency and for different plasma temperature. $\alpha\geq0$ ($\alpha<0$) for the LCP (RCP). Baseline plasma parameters are given in Table~\ref{Tab:params}. }
\label{Fig:Fig2}
\end{center}
\end{figure}

In contrast, losses in a rotating plasma are small. In a fully ionised plasma away from resonances, losses result from Coulomb collisions~\cite{Fleishman2013}. For $\zeta=\omega/\omega_{ci}\ll1$, to lowest order in $\zeta$, the imaginary part of the ${n_{rcp/lcp}}^2$ in Eq.~(\ref{Eq:index_general}) comes from $\Im(\bar{\chi}_{\parallel}) = -\Re(\bar{\chi}_{\parallel})\nu_{ei}/\omega$ with $\nu_{ei}$ the electron-ion collision frequency~\cite{GueroultSupp}. Taylor expanding Eq.~(\ref{Eq:index_general}) for complex valued susceptibilities, the wave index for LCP and RCP wave at the cutoff gives
\begin{equation}
n_l(\omega_c) \sim (1+i)\sqrt{\tau} \quad \textrm{and} \quad n_r(\omega_c)  \sim \sqrt{2}\left(1-i\frac{\tau}{2}\right)
\label{Eq:complex_index}
\end{equation}
where we assumed $\tau=\nu_{ei}/(2\omega_c)\ll1$. Taking $\nu_{ei}$ as the standard Lorentz collision frequency, one finds that this last condition is verified for the baseline parameters given in Table.~\ref{Tab:params} as long as the electron temperature $\textrm{T}_e\geq20$~eV. Solving numerically Eq.~(\ref{Eq:index_general}) for the absorption coefficient $\alpha(\omega) = 2 \omega \Im[n(\omega)]/c$ shows, as plotted in Fig.~\ref{Fig:Fig2}, that this condition further ensures $\alpha\ll1$~m$^{-1}$. Note that $\alpha$ and thus $T$ decrease with frequency above the cutoff, which is in contrast with the rapid drop of transmission observed in permanent magnets~\cite{Shalaby2013} and graphene~\cite{Qin2018}.  Note also from Eq.~(\ref{Eq:complex_index}) that $\Im(n)$ at the cutoff differs for LCP and RCP. This difference will lead to circular dichroism, in addition to the circular birefringence induced by the difference in the real part of wave indexes. However, Fig.~\ref{Fig:Fig2} shows that this will only occur in a narrow frequency band over the cut-off and will have limited effect. Indeed the ellipticity per unit length $\psi = \Im[\Delta n(\omega)]\omega/(2c) = \Delta \alpha/4$ is at least 10 times smaller than $\delta$ for $T_e = 20$~eV, and the ratio $\psi/\delta$ decreases as ${T_e}^{-3/4}$. 

\paragraph*{High figure of merit.} A commonly used figure of merit (\emph{fom}) to compare the NR performances of different media is $\varsigma = \theta\sqrt{T}$ with $\theta$ is the rotation angle in rad, which combines information on specific rotatory power and losses. From Eq.~(\ref{Eq:rotatory_power}), the \emph{fom} in a rotating plasma at the cutoff writes
\begin{equation}
\varsigma(\omega=\omega_c)\sim\frac{\omega_c}{\sqrt{2}c}d\exp\left[-\frac{\omega_c\sqrt{\tau}d}{c}\right].
\label{Eq:fom_plasma}
\end{equation} 

\begin{figure}
\begin{center}
\includegraphics[]{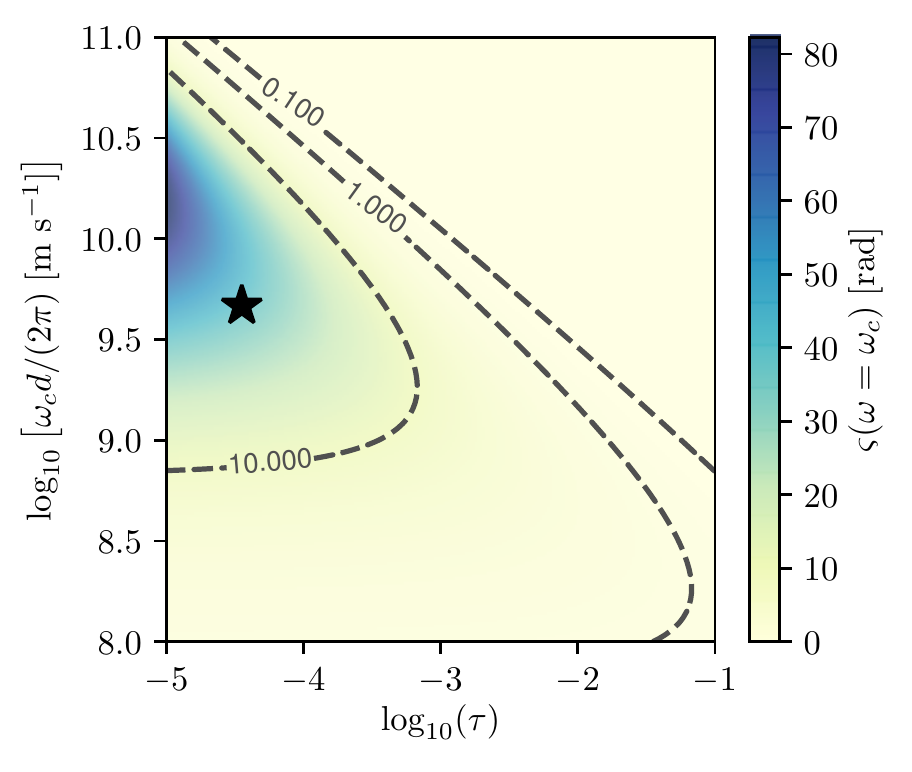}
\caption{Figure of merit $\varsigma$ at the cutoff as a function of the product $\omega_c d$ of the cutoff frequency times the length and of the collision frequency normalized by the cutoff frequency $\tau=\nu_{ei}/(2\omega_c)$. The star symbol ($\varsigma\sim40$) corresponds to THz operation for the plasma parameters (${n_e}^{T}$,${B_0}^T$,$\Omega^T$) given in Table~\ref{Tab:params2}, T$_e = 5$~eV and $d = 5$~mm. }
\label{Fig:Fig3}
\end{center}
\end{figure}

Eq.~(\ref{Eq:fom_plasma}) underlines two key advantages of rotating plasmas. First, since $\alpha(\omega)$ peaks at the cutoff, both the polarization angle $\theta$ and transmission $T$ increase with $\omega_c$, favouring high frequency operation. Second, the thickness $d$ in a plasma is a free parameter which can be chosen to maximize $\varsigma$. This is illustrated in Fig.~\ref{Fig:Fig3} where the \emph{fom} is plotted as a function of the losses parameter $\tau$ and the product of the cutoff frequency times propagation length $d$. One finds that $\varsigma$ indeed first grows with $d$ until reaching a maximum past which it starts decreasing.

\paragraph*{Opportunities for THz operation. } Since $\omega_c\sim[{\omega_{pe}}^2\Omega]^{1/3}$, high frequency operation requires fast rotation and dense plasmas. While kHz rotation has been achieved in plasmas as a result of cross-field drift~\cite{Ivanov2013} or cross-field current~\cite{Flanagan2020} and MHz rotation could possibly be produced by ionizing a rapidly rotating gas~\cite{Steinitz2012,Kaganovich2020}, it is likely insufficient for operation beyond a few GHz. Quantitatively, THz operation requires $\Omega^{T}/(2\pi)\geq10^{11}$~Hz for a plasma density ${n_e}^{T}=10^{23}$~m$^{-3}$ which appears impractical in a neutral plasma. On the other hand, it happens that $\Omega^{T}$ is precisely the angular rotation frequency of a non-neutral plasma with $n_e^{T}\gg {n_i}^{T}$. Due to its space-charge radial electric field, a non-neutral plasma indeed naturally rotates at $\Omega={\omega_{pe}}^2/(2\omega_{ce})$~\cite{Davidson2001}.


Although the detailed properties of polarization drag differ slightly in the case of a non-neutral plasma, the enhancement of polarization rotation above the cutoff angular frequency $\omega_c\sim[{\omega_{pe}}^2\Omega]^{1/3}$ holds in an electron dominated plasma~\cite{GueroultSupp}. Owing to the frequency dependence of $\delta_m$ shown in Eq.~(\ref{Eq:rotatory_power}), THz operation for the parameters given in Table~\ref{Tab:params2} would then yield a giant polarization drag with $\delta\geq10^4$~rad~m$^{-1}$. In these conditions propagation over a distance $d_0=0.1$~mm would produce a $\pi/4$ polarization rotation. 

While such specific rotatory powers remain lower than the state-of-the-art at THz frequencies on epitaxial HgTe thin films~\cite{Shuvaev2011} or in graphene~\cite{Crassee2010}, the low losses exhibited by plasmas translate into significantly larger \emph{fom}. This is particularly true at high frequency (\emph{i.~e.} high $\omega_c$) since $\varsigma$ depends on $\tau=\nu_{ei}/(2\omega_c)$ and in a non-neutral plasma since $\nu_{ei}$ is proportional to $n_i$. For the parameters given in Table~\ref{Tab:params2} a $5$~mm long plasma yields $\theta\sim90$~rad, and an electron temperature $\textrm{T}_e\sim20~$eV is then enough to reach $\varsigma\sim45$~rad (star symbol in Fig.~\ref{Fig:Fig3}) for $n_e/n_i=100$. This is already significantly larger than the best values $\varsigma\leq1$~rad achieved in permanent magnets~\cite{Shalaby2013} and graphene metasurfaces~\cite{Qin2018}, and higher $\varsigma$ up to $\theta$ could be produced in plasma at higher electron temperatures or higher density asymmetry ratios $n_e/n_i$.

Finally, while producing such plasmas remains to be demonstrated experimentally, note that it is at least energetically feasible. Considering for simplicity a spherically symmetric volume, a non-neutral plasma of radius $a=1$~mm and density ${n_e}^{T}$ has a net charge $Q \sim 10^{-4}~$C, and the energy required to excavate this charge  from an originally neutral plasma is $\mathcal{E}\propto Q^2/a\sim10^4$~J. The multi kJ lasers available on high energy density plasma (HEDP) platforms~\cite{Waxer2005,Haynam2007,Blanchot2008} could thus in principle produce non-neutral plasmas in volumes larger than ${d_0}^3$ ($d_0\sim a/10$) as needed for THz operation.

\begin{table}
\begin{center}
\caption{Plasma parameters for THz operation and corresponding circular birefringence properties}
\label{Tab:params2}
\begin{tabular}{c c c }
\hline
\hline
Electron density & ${n_e}^{T}$ & $10^{23}$~m$^{-3}$\\
Magnetic field & ${B_0}^{T}$ & $10^3$~T\\
Rotation frequency & $\Omega^{T}$ & $10^{11}$~Hz\vspace{0.1cm}\\
Cutoff frequency & $\omega_c$ & $1$~THz\\
Specific rotatory power & $\delta$ & $10^4$~rad~m$^{-1}$\\
\hline
\hline
\end{tabular}
\end{center}
\end{table}

\paragraph*{Conclusions. } We identified that the mechanical rotation of a magnetized plasma enhances circular birefringence. An additional cutoff for propagation along the magnetic field is produced, so that polarization drag can be orders of magnitude larger than the classical magneto-optical (Faraday) rotation at GHz frequency. This effect, akin to that previously observed using slow light in a rotating solid dielectric~\cite{Franke-Arnold2011}, can be explained as the result of a very large effective group index. 

We also demonstrated that both the cutoff frequency and the specific rotatory power can be controlled through the plasma angular rotation frequency and the plasma density. In considering collisional damping in a plasma, we further showed that the weak collisional absorption in a plasma compared to alternative non-reciprocal materials is expected to lead to a favorable mechanico-optical figure of merit. 

Finally, building on our finding that the specific rotatory power grows linearly with the cutoff frequency, we inferred that a rotating magnetized plasma could offer unprecedented non-reciprocal capability in the THz regime. We suggest that THz operation could be achieved by leveraging the very large rotation naturally arising in a dense non-neutral plasma. 

Hence, the present Letter reveals the unique non-reciprocal properties of rotating magnetized plasmas and paves the way for demonstrating experimentally polarization drag effects in a plasma. These results also provide the basis for the development of tuneable and reconfigurable non-reciprocal systems using rotating plasmas. In uniquely enabling real time tuning to adapt to a change in wave frequency $\omega$, a rotating plasma could for instance provide the foundations for a compact wavelength-agile THz isolator.

\begin{acknowledgments}
This work was supported, in part, by grant No. DOE NNSA DE-NA0003871.
\end{acknowledgments}


%

\end{document}